\documentclass[%
 reprint,
 amsmath,amssymb
]{revtex4-2}

\usepackage{graphicx}
\usepackage{dcolumn}
\usepackage{bm}

\begin{document}

\title{Dirac Points Embedded in the Continuum}

\author{Pilar Pujol-Closa}
\author{Lluis Torner}
\author{David Artigas}
\email{david.artigas@icfo.eu}
 \affiliation{ICFO - Institut de Ciencies Fotoniques, The Barcelona Institute of Science and Technology}%
\affiliation{Department of Signal Theory and Communications, Universitat Polit\`ecnica de Catalunya
}


\begin{abstract}

\noindent Dirac points (DP) in Hermitian systems play a key role in topological phenomena. Their existence in non-Hermitian systems is then desirable, but the addition of loss or gain transforms DPs into pairs of Exceptional Points (EPs) joined by a Fermi arc, which exhibit interesting but different properties. When the transition to a non-Hermitian system results from the opening of a radiation channel, the system can also support bound states in the continuum (BICs), which are non-radiative resonant states that appear within the band of radiation states. We theoretically show that simultaneous band-crossing of two BICs can prevent the formation of EPs and Fermi arcs, resulting in genuine Hermitian DPs, which are nonetheless embedded in the continuum of radiation states. Dirac points embedded in the continuum (DECs) are a new topological entity that combines the rich physics associated with DPs with the ideal resonant properties of BICs in non-Hermitian systems.

\end{abstract}

\maketitle

\noindent Degeneracy between energy bands has raised interest since the beginning of quantum mechanics \cite{Herring}. A Dirac point (DP) in a Hermitian system occurs at the intersection in a point of two dispersion bands, exhibiting a linear slope that forms a conical surface \cite{Haldane, graphene, Beyond-Dirac}. The result is a degenerate state with identical eigenvalues and two orthogonal eigenstates. Recently, the topological concepts and applications of Hermitian systems are being exported to non-Hermitian systems \cite{non-H-topo,non-H-band-theory, non-H-edge-states}, in particular, DPs \cite{non-H-DP,special-EP,complex-DP}. However, when the system becomes non-Hermitian, the eigenstates cease to be orthogonal and the eigenvalues become complex, with the imaginary part being related to losses. In this transformation, the DP gives rise to a pair of Exceptional points (EP) connected by a Fermi arc \cite{rings-EPs,ozdemir}. Just at the EP, the two bands coalesce and the Hamiltonian is described by a non-diagonal Jordan matrix with identical complex eigenvalues and identical eigenstates \cite{non-hermitian}. The band crossing is produced at the Fermi arc, forming the two halves of a Riemann surface. Non-orthogonal eigenstates in non-Hermitian systems result in non-trivial dynamics \cite{non-hermitian, alu-review}, including asymmetric mode switching \cite{Doppler,yoon}, topological half-charges in polarization states \cite{Zhou-sci}, high sensitive measurements \cite{hodaei,chen-nat17} or directional lasing and chiral modes \cite{pengPNAS}. 

Open systems are a particular kind of non-Hermitian systems. The eigenstates in these systems are resonances coupled to the continuum, resulting in energy radiation. Cancellation of the radiation channels by different mechanisms, including symmetry protection or destructive interference, results in Bound states in the continuum (BICs). BICs appear in the parameter space corresponding to the radiation continuum as confined eigenstates inserted within the dispersion band of complex eigenvalues. However, the eigenvalue in a BIC is real, thus the radiation channel is suppressed, resulting in an infinite lifetime or propagation distance. Although BICs were proposed in quantum mechanics \cite{Neuman1929, Stillinger1975}, they are a general wave phenomena \cite{Hsu2016} and have been found in a variety of physical settings including acoustic \cite{Parker1966}, quantum \cite{Kim1999}, and, prominently photonic \cite{Marinica2008, Bulgakov2008, Plotnik2011, Hsu2013, Monticone2014} platforms.

A unique situation occurs when BICs interact with EPs, resulting in BICs flipping position with EPs \cite{Kikkawa_2020}, exchanging of dispersion band at the Fermi arc \cite{Cerjan2021}, or obtaining ultra-low loss EPs in a dual-BIC scheme \cite{shi2022}. Here, we explore a new situation where two BICs meet and interchange bands by crossing the Fermi arc. As the eigenvalues of the two BICs are purely real, the system is locally Hermitian at the crossing point. Interestingly, when the BICs resonances are broad enough, the EPs cease to exist and a gap is opened along the Fermi arc, except at the point where the two BICs cross. At this point, the two dispersion bands join with a conical and linear slope, the eigenstates are orthogonal, and the eigenvalues are real and degenerated, thus resembling a DP. However, this finding differs from previous approaches, where the DP itself is non-Hermitian \cite{non-H-DP,special-EP,complex-DP}, in that radiation losses are totally canceled at the DP, while it is surrounded by non-orthogonal eigenstates with complex eigenvalues. This is therefore a new paradigm where a Hermitian DP is embedded within the non-Hermitian continuum. A Dirac point embedded in the continuum (DEC) is a new topological entity that can retain all the physics related to DPs (topologically protected edge states, back-scattering immune transport, \cite{Qhallgraph, edgephoton, RevModPhys}) in addition to the resonant properties associated with BICs \cite{mergingBICs}, and potentially opens the door to a plethora of new phenomena yet to be discovered.

In this paper, we use a simple model based on a generic two-level system to demonstrate the formation of DECs. The concept is then corroborated in a more complex photonic system based on hyperbolic waveguides \cite{hypPRL-03, nanowiresSC08, bulkhyp, Takayama:19} with anisotropic background \cite{Gomis-Bresco2017, GomisDPEP}, showing a perfect agreement.

The two-level system models the coupling between two resonances with amplitudes E$_1$ and E$_2$, which evolve as a function of a parameter, in our case the propagation distance $z$. Then, considering the mode propagation constant or momentum, $\kappa_j$, the attenuation constant associated to the radiation channel, $\alpha_j$, and the coupling between the two resonances, $q$, the dynamics of the system is described by the Hamiltonian $H$ \citep{Doppler, alu-review}:

\begin{equation}
    i \dfrac{d}{dz} \begin{bmatrix} E_1 \\ E_2 \end{bmatrix} = H \begin{bmatrix} E_1 \\ E_2 \end{bmatrix} = \begin{bmatrix} \kappa_1-i \alpha_1 & q \\ q & \kappa_2 - i \alpha_2  \end{bmatrix} \begin{bmatrix} E_1 \\ E_2 \end{bmatrix}, 
    \label{Hamiltonian}
\end{equation}

\noindent  The system supports harmonic solutions of the type $E_j=a_j e^{i \beta_j z}$, where the eigenvalue $\beta$ can be described in terms of the momentum mismatch, $\Delta \kappa = \kappa_2 - \kappa_1$, and the attenuation mismatch, $\Delta \alpha = \alpha_2 - \alpha_1$, as:

\begin{equation}
    \beta =  \left( \kappa_{av} - i \alpha_{av} \pm \dfrac{1}{2}\sqrt{4q^2 + \left(\Delta \kappa - i \Delta \alpha \right) ^2 } \right). 
    \label{eigenvalues}
\end{equation}

\begin{figure*}[ht]
\centering
\includegraphics[width=0.7\linewidth]{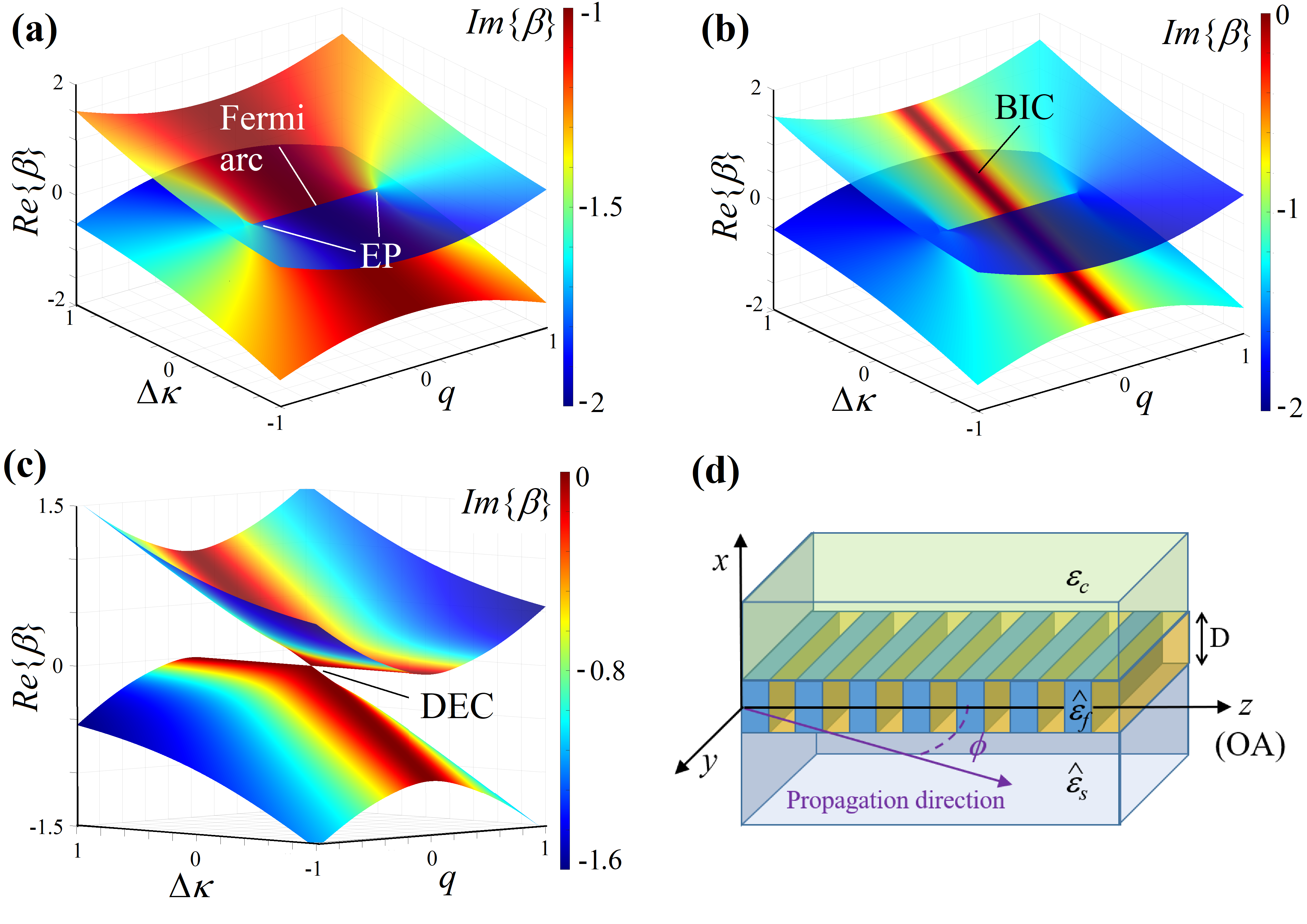}
\caption{Eigenvalues bands of a two-level system in terms of the coupling $q$ and the phase mismatch $\Delta \kappa$, with $\alpha_1=1 $ and $\alpha_2=2$. (a) Losses are constant for any value of $q$, and $\Delta \kappa$. (b), The E$_1$ level contains a BIC line with a width resonance $w_1=0.1$, the result is the BIC line at $q=0$ passing through the Fermi arc. (c), Both resonances contain a BIC line at $q=0$ with a width resonance $w_1=0.4$, which results in a DEC. (d) Hyperbolic photonic structure composed of an isotropic cladding, a negative birefringent substrate, and a Type II hyperbolic film. The OA is oriented along the $z$ direction. The angle $\phi$ indicates the propagation direction.}
\label{Fig1}
\end{figure*}

\noindent Here, $\kappa_{av} =( \kappa_2 + \kappa_1$)/2 and $\alpha_{av} = (\alpha_2 + \alpha_1$)/2 are the averaged momentum and attenuation constant, respectively. Fig. \ref{Fig1}(a) shows the typical eigenvalue bands of the system as a function of the coupling $q$ and the momentum mismatch $\Delta \kappa$. The EP appears at $\Delta \kappa =0$ and $q= \Delta \alpha/2$, with eigenvalue $\beta = \kappa_{av} - i \alpha_{av}$. At the Fermi arc ($\Delta \kappa =0$), both eigenvalues intersect with equal $Re(\beta)$ but different $Im \left(\beta \right)$. This results in the line with different losses joining the two EPs in Fig. \ref{Fig1}a. When $\Delta \alpha=0$, the two EPs collapse at $\Delta \kappa =0$, $q= 0$, into a single EP, or complex DP, with one complex eigenstate \cite{complex-DP}. Consequently, an attenuation mismatch is needed for the Fermi arc to exist.

Contrary to absorption losses, radiation losses are dictated by the structural and geometrical parameters describing an open system. In particular, losses can be canceled due to the presence of a BIC, which usually appears as isolated points within the leaky dispersion band. However, in systems with a single radiation channel, BICs can form continuous lines of real eigenvalues inserted within dispersion bands of complex eigenvalues. The latter situation has been predicted in anisotropic waveguides \cite{Gomis-Bresco2017, Mukherjee2018}, photonic crystal slabs with environment design \cite{Cerjan2021}, and appears naturally in surface BICs \cite{Mukherjee-dya}. Taking into account that radiation losses are related to the resonant quality factor by $\alpha_i \propto Q^{-1}$ and following the quadratic scaling rule of $Q$ in terms of the system parameters \cite{mergingBICs}, a line of BICs in level $i$ can conveniently be modeled in the Hamiltonian (\ref{Hamiltonian}) by an absorption:
\begin{equation}
    \alpha_i=\alpha_{0i}q^2/(q^2+w^2_i),
    \label{alpha}
\end{equation}
\noindent where $w_i$ is the width of the BIC resonance in level $i$, and $\alpha_{0i}$ is the radiation loss far away from the BIC ($q >> w_i$).

The situation described in Ref. \cite{Cerjan2021} can be simulated by considering a single BIC in the $E_1$ resonance and choosing $w_i=0.1$ to minimally affect the position of the EPs. The result in Fig. \ref{Fig1}b is a BIC line that crosses the Fermi arc, exchanging dispersion band. The presence of the BIC at $q=0$ results in $Im \{\beta_1 \}=0$, while the positions of the EPs have not changed perceptibly. BICs with a broader resonance can affect the EPs position, but the EPs do not disappear.

A more interesting situation occurs when both resonances present a BIC. By tuning the system parameters, the two BICs can be made to coincide at the Fermi arc where they exchange band. Figure \ref{Fig1}c shows the situation for two lines of BICs with $w_1=w_2=0.4$. A gap between the two bands is opened all along the Fermi arc except at the crossing between the two BICs. Examination of the point where BICs cross shows that two orthogonal eigenstates coexist, the eigenvalues are identical and real, and the slope of the eigenvalue band at this point is linear showing a conical surface. Thus, this point is not the coalescence of two bands in a single EP or complex DP \cite{complex-DP}, as the eigenvalues are real. Neither is a non-Hermitian DP where gain a loses are balanced \cite{non-H-DP}, as in our case the radiation losses at the point where the two BICs cross is zero and thus the system is locally Hermitian. However, the point is surrounded by complex eigenvalues, which are non-Hermitian. Therefore, its properties correspond to a DP, which however is located in the space of parameters that corresponds to the radiation continuum. In other words, it is a Dirac point embedded in the continuum (DEC). DECs existence only requires that the BICs resonances are broad enough to avoid the existence of EPs and are also possible when the crossing BICs are points in the dispersion band. 

To show DECs existence in a specific system, we consider a structure in Fig. \ref{Fig1}d based on a planar waveguide made of a Type II hyperbolic film (or core), a birefringent substrate with elliptical dispersion, and an isotropic cladding. Each media in the waveguide is characterized by a diagonal permittivity tensor given by $\hat{\epsilon}= diag \left (\epsilon_o, \epsilon_o, \epsilon_e  \right )$. The hyperbolic film, with a thickness $D$, is characterized by positive extraordinary and negative ordinary dielectric constants ($\varepsilon_{ef}>1$, $\varepsilon_{of}<0$), i.e., Type II hyperbolic dispersion \cite{Takayama:19}. The substrate shows negative birefringence, with  $\varepsilon_{os}>\varepsilon_{es}>1$. The refractive index of the isotropic dielectric cladding is $\varepsilon_c \geq 1$. We consider the film and substrate optical axes (OAs) to be aligned and lay parallel to the waveguide interfaces. The angle $\phi$ shows the propagation direction with respect to the OA (Fig. \ref{Fig1}d).

We analyze the structure using a modified transfer matrix method, where part of the code includes analytical routines to improve accuracy \cite{Gomis-Bresco2017, Mukherjee2018, Pujol-Closa2021}. The result provides the mode field amplitudes and eigenvalues, i.e., the effective index $N$, which is related to the eigenvalue or propagation constant, $\beta$, as $N=\beta/k_0$, with free space wavenumber $k_0=2\pi/\lambda$ and free space wavelength $\lambda$. Being a birefringent structure, $N$ changes with the propagation direction $\phi$. In the case of negative birefringent substrates, the waveguide supports guided modes when $N > n_{os}$ and leaky modes when $n_{es}< N < n_{os}$. Leaky modes are improper complex solutions of the eigenmode equation where the imaginary part of $N$ gives a good approximation to the actual radiation losses and the eigenvector properly describes the field at the vicinity of the film. They form leaky branches, which are the equivalent of resonant bands in generalized non-Hermitian radiating systems. The waveguide under study can support different types of modes with hybrid transverse electric and magnetic (TE/TM) polarization. We are interested in the modes existing near $\phi = 90^\circ$, which consist of a finite set of TE-dominant normally ordered hyperbolic (TENH) modes and a TM-dominant (TMd) plasmon \cite{Pujol-Closa2021}. As $D/\lambda$ increases, TENH modes grow above the cutoff, first as a leaky mode to become guided when $N>n_{os}$. The TMd plasmon depends primordially on the film/cladding interface properties and can be designed to be leaky in the region of interest. In addition, the resulting leaky branches support interference BICs (INT-BICs), and polarization separable BICs (PS-BICs), which are the equivalent in other systems to accidental and symmetry-protected BICs, respectively \cite{Gomis-Bresco2017}. INT-BICs feature a hybrid TE/TM polarization with a propagation direction $\phi$ that can be parameter tuned in terms of the waveguide parameters \cite{Mukherjee2018, Mukherjee2019}. In contrast, PS-BICs are either pure TE or TM polarized and appear at the propagation direction orthogonal to the OA, $\phi=90^\circ$ \cite{Gomis-Bresco2017}. 

Without loss of generality, the results were obtained using the following values for the relative permittivity: $\varepsilon_c=1$ for the cladding, $\varepsilon_{ef}= 1.75^2$ and $\varepsilon_{of}=-1.77$ for the hyperbolic film, and $\varepsilon_{es}=1.25^2$ and $\varepsilon_{os}=4$ for the substrate. Note that the ordinary wave in the substrate is the radiation channel as it features the higher $\varepsilon_r$ of the cladding and substrate. Non-local effects and material absorption are not considered \cite{Lavrinenko}, which is consistent with experimental structures recently demonstrated using a film of hyperbolic natural materials \cite{hyperwave}. 

\begin{figure*}[ht]
\centering
\includegraphics[width=0.7\linewidth]{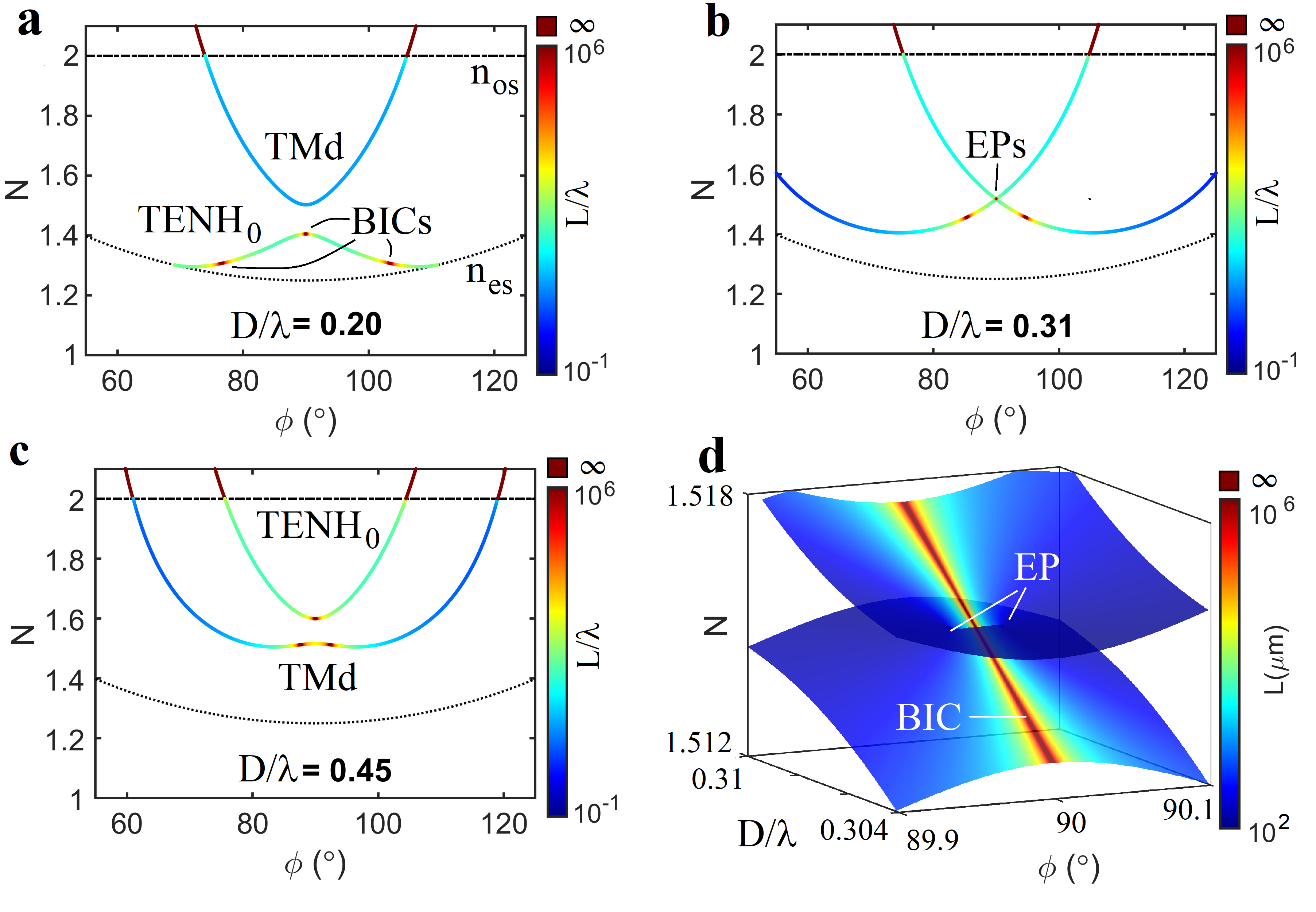}
\caption{(a), (b), and (c), angular dispersion diagram showing the effective index $N$ in terms of the propagation direction $\phi$ for different electrical thicknesses $D/\lambda$ (indicated in each panel). The line color in every branch indicates the normalized mode decay length $L=1/\beta$. The black dot and dot-dashed lines indicate the substrate extraordinary, $n_{es}$, and ordinary, $n_{os}$, refractive indices, respectively. TMd stands for the TM-dominant plasmon and TENH$_i$ for the TE-dominant, normally-ordered hyperbolic mode with order index $i$. (d) Effective index bands in terms of $\phi$ and $D/\lambda$ showing the EPs and one PS-BIC crossing the Fermi arc.}
\label{dispEP}
\end{figure*}

The nature of BICs and their relationship with EPs and DECs can be unveiled by analyzing the angular dispersion diagrams, i.e., the effective index for the modes supported by the structure in terms of the propagation direction $\phi$. We do that for different electrical thicknesses $D/\lambda$, considering $\lambda$ fixed. Experimentally, in low-dispersive materials, varying the operational wavelength for a fixed $D$ would yield similar qualitative results. The two mode dispersion curves for $D/\lambda=0.20$ in Fig. \ref{dispEP}a correspond to a TMd plasmon (higher $N$), which at $\phi=90^\circ$ is pure TM, becoming hybrid TM dominant all along the branch. At $74^\circ < \phi < 106^\circ $ the TMd plasmon is leaky ($N<n_c=2$), while beyond this range is guided ($N>n_c=2$). The second mode is the fundamental TENH$_0$, which is leaky in all the range of existence except at the BICs. This branch supports three BICs: a PS-BIC at $\phi=90^\circ$ with TE polarization and an INT-BIC at each side. As $D/\lambda$ increases, the TENH$_0$ branch approaches the TMd-plasmon branch, until at $D/\lambda \approx 0.31$ the two branches seem to intersect at $\phi=90^\circ$ (see Fig. \ref{dispEP}b). If $D/\lambda$ increases further, the two branches depart apart, as shown by Fig. \ref{dispEP}c for $D/\lambda=0.45$. However, two important effects should be highlighted. First, the branches at $\phi=90^\circ$ exchange polarization, so that at $d/\lambda=0.45$ the lower branch corresponds to the TMd-plasmon and the upper branch is the TENH$_0$ mode. Second, the PS-BIC crosses to the upper branch, while the INT-BICs remain at the lower branch, approaching each other as $D/\lambda$ increases. A close examination of this transition in Fig. \ref{dispEP}d unveils the existence of two EPs and a Fermi arc at $D/\lambda=0.307$, and the crossing of the PS-BIC through the Fermi arc, exchanging branch. The result locally resembles Fig. \ref{Fig1}b, with $\phi$ and $D/\lambda$ playing the role of the coupling between states $q$ and the propagation constant mismatch $\Delta \kappa$, respectively. Note that at the point $\phi=90^\circ$ in the Fermi arc, there are two solutions for $N$, one of them is real (the BIC) and the other complex (the TMd plasmon), therefore the system is non-Hermitian for all points in Fig. \ref{dispEP}d.

\begin{figure*}[ht]
\centering
\includegraphics[width=0.7\linewidth]{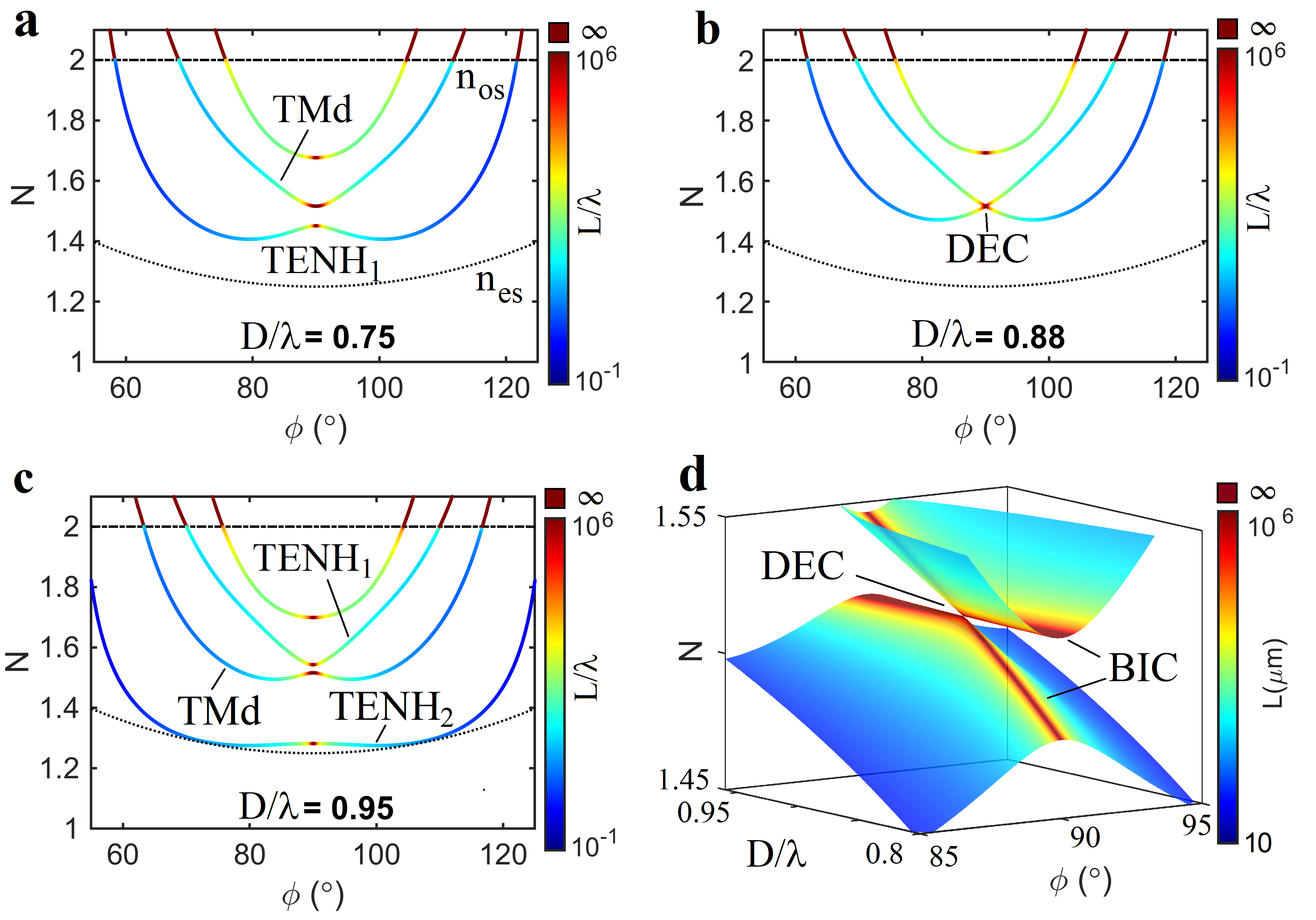}
\caption{(a), (b), and (c) same as Fig. \ref{dispEP} but for higher electrical thicknesses $D/\lambda$, as indicated in each panel. {\bf d,} Effective index bands in terms of $\phi$ and $D/\lambda$ showing a Dirac Point embedded in the continuum (DEC).}
\label{dispDEC}
\end{figure*}

As the electrical thickness $D/\lambda$ increases further, the two INT-BICs in the TMd plasmon branch finally joints at $\phi=90^\circ$, resulting in a wide area with low losses \cite{mergingBICs}. In addition, a new TENH$_1$ mode appears from the cutoff featuring a PS-BIC (Fig. \ref{dispDEC}a). Near $D/\lambda \approx 0.88$, the TMd plasmon and TENH$_1$ bands intersect, with both branches having a BIC at the intersection point $\phi=90^\circ$ (Fig. \ref{dispDEC}b). As $D/\lambda$ increases, both branches again split apart (Fig. \ref{dispDEC}c). By examining the electromagnetic field at each branch, we observe that the branches have exchanged polarization again, so that the second branch in Fig. \ref{dispDEC}c now is the TENH$_1$ mode, while the third branch is the TMd plasmon. We can also observe that the wide low loss area resulting after the BICs merging is now in the third branch, confirming the BIC exchange of branches. However, the underlying phenomena, in this case, is different from Fig. \ref{dispEP}, as now two BICs are crossing at the same point, which is the necessary condition to obtain a DEC. A close examination by plotting the two effective index bands, $N$, in terms of $\phi$ and $D/\lambda$ (Fig. \ref{dispDEC}d) shows that in this case, the two bands cross at a single point, $\phi=90^\circ$ and $D/\lambda=0.8804$, the two electromagnetic fields at this point are orthogonal and, as both of them are non-radiating BICs, both eigenvalues $N$ are real, thus the system is locally Hermitian. In addition, the two bands show a conical slope, resulting in a DEC which greatly resembles the one described by the much simpler two-level model in Fig. \ref{Fig1}c. Finally, note that for $D/\lambda= 0.95$, a new branch of TENH$_2$ mode appears above the cutoff in Fig. \ref{dispDEC}c. This mode has a PS-BIC at $\phi=90^\circ$ and as $D/\lambda$ increases further, the branch will cross the TMd-plasmon branch, resulting in a new DEC. This process will be repeated for all the new TENH$_n$ branches appearing as $D/\lambda$ is increased. 

Photonic waveguides with a hyperbolic film are a convenient setting to put forward the DEC concept. Hermitian systems based on planar waveguides can present DPs that transform into EPs when a radiation channel is opened \cite{GomisDPEP}. The  much richer variety of modes in the hyperbolic film provides the leaky TMd plasmon, which can support INT-BICs. This being a surface wave, $N$ is not affected by a $D/\lambda$ increase, allowing its interaction with the consecutive leaky TENH branches. This interaction can originate either pairs of EPs united by Fermi arcs, which can be crossed by one BIC, or DECs. We however advance DECs existence in other photonic arrangements that support BICs, such as coupled waveguides, photonic crystals, or resonant structures, as well as in acoustic or quantum systems.

DECs are a new topological entity enriching the fields of topological physics. Similarly to standard DPs, two conical eigenvalue surfaces intersect at the DEC with a single real eigenvalue and two coexisting orthogonal eigenstates. However, they originate due to the interaction of two BICs, which prevent the existence of EPs and open a gap all along the Fermi arc except at the DEC. When compared with standard BICs, a DEC offers a unique characteristic, as it is a degenerate resonance that provides infinite propagation distances or lifetimes for two different orthogonal states. In addition, although DECs show the same features as DPs in a Hermitian system, they are surrounded by radiation states. In other words, unlike other DPs in non-Hermitian systems that requires anti-PT symmetry \cite{non-H-DP}, DECs are surrounded by states where the PT-symmetry does not hold. Consequently, all the physical properties of DP in Hermitian systems need to be reevaluated in the framework of DECs, which may lead to novel forms of electromagnetic and quantum manipulation of wave states.

\noindent {\bf Acknowledgments:} This work was partially supported by the Ministerio de Econom\'ia y Competitividad, Grants CEX2019-000910-S and PGC2018-097035-B-I00 funded by MCIN/AEI/10.13039/501100011033/FEDER, Fundaci\'o Cellex, Fundaci\'o Mir-Puig, and Generalitat de Catalunya (CERCA). 

\bibliography{decs}

\begin{thebibliography}{49}%
\makeatletter
\providecommand \@ifxundefined [1]{%
 \@ifx{#1\undefined}
}%
\providecommand \@ifnum [1]{%
 \ifnum #1\expandafter \@firstoftwo
 \else \expandafter \@secondoftwo
 \fi
}%
\providecommand \@ifx [1]{%
 \ifx #1\expandafter \@firstoftwo
 \else \expandafter \@secondoftwo
 \fi
}%
\providecommand \natexlab [1]{#1}%
\providecommand \enquote  [1]{``#1''}%
\providecommand \bibnamefont  [1]{#1}%
\providecommand \bibfnamefont [1]{#1}%
\providecommand \citenamefont [1]{#1}%
\providecommand \href@noop [0]{\@secondoftwo}%
\providecommand \href [0]{\begingroup \@sanitize@url \@href}%
\providecommand \@href[1]{\@@startlink{#1}\@@href}%
\providecommand \@@href[1]{\endgroup#1\@@endlink}%
\providecommand \@sanitize@url [0]{\catcode `\\12\catcode `\$12\catcode
  `\&12\catcode `\#12\catcode `\^12\catcode `\_12\catcode `\%12\relax}%
\providecommand \@@startlink[1]{}%
\providecommand \@@endlink[0]{}%
\providecommand \url  [0]{\begingroup\@sanitize@url \@url }%
\providecommand \@url [1]{\endgroup\@href {#1}{\urlprefix }}%
\providecommand \urlprefix  [0]{URL }%
\providecommand \Eprint [0]{\href }%
\providecommand \doibase [0]{https://doi.org/}%
\providecommand \selectlanguage [0]{\@gobble}%
\providecommand \bibinfo  [0]{\@secondoftwo}%
\providecommand \bibfield  [0]{\@secondoftwo}%
\providecommand \translation [1]{[#1]}%
\providecommand \BibitemOpen [0]{}%
\providecommand \bibitemStop [0]{}%
\providecommand \bibitemNoStop [0]{.\EOS\space}%
\providecommand \EOS [0]{\spacefactor3000\relax}%
\providecommand \BibitemShut  [1]{\csname bibitem#1\endcsname}%
\let\auto@bib@innerbib\@empty
\bibitem [{\citenamefont {Herring}(1937)}]{Herring}%
  \BibitemOpen
  \bibfield  {author} {\bibinfo {author} {\bibfnamefont {C.}~\bibnamefont
  {Herring}},\ }\bibfield  {title} {\bibinfo {title} {Accidental degeneracy in
  the energy bands of crystals},\ }\href
  {https://doi.org/10.1103/PhysRev.52.365} {\bibfield  {journal} {\bibinfo
  {journal} {Phys. Rev.}\ }\textbf {\bibinfo {volume} {52}},\ \bibinfo {pages}
  {365} (\bibinfo {year} {1937})}\BibitemShut {NoStop}%
\bibitem [{\citenamefont {Haldane}(1988)}]{Haldane}%
  \BibitemOpen
  \bibfield  {author} {\bibinfo {author} {\bibfnamefont {F.~D.~M.}\
  \bibnamefont {Haldane}},\ }\bibfield  {title} {\bibinfo {title} {Model for a
  quantum hall effect without landau levels: Condensed-matter realization of
  the "parity anomaly"},\ }\href {https://doi.org/10.1103/PhysRevLett.61.2015}
  {\bibfield  {journal} {\bibinfo  {journal} {Phys. Rev. Lett.}\ }\textbf
  {\bibinfo {volume} {61}},\ \bibinfo {pages} {2015} (\bibinfo {year}
  {1988})}\BibitemShut {NoStop}%
\bibitem [{\citenamefont {Castro~Neto}\ \emph {et~al.}(2009)\citenamefont
  {Castro~Neto}, \citenamefont {Guinea}, \citenamefont {Peres}, \citenamefont
  {Novoselov},\ and\ \citenamefont {Geim}}]{graphene}%
  \BibitemOpen
  \bibfield  {author} {\bibinfo {author} {\bibfnamefont {A.~H.}\ \bibnamefont
  {Castro~Neto}}, \bibinfo {author} {\bibfnamefont {F.}~\bibnamefont {Guinea}},
  \bibinfo {author} {\bibfnamefont {N.~M.~R.}\ \bibnamefont {Peres}}, \bibinfo
  {author} {\bibfnamefont {K.~S.}\ \bibnamefont {Novoselov}},\ and\ \bibinfo
  {author} {\bibfnamefont {A.~K.}\ \bibnamefont {Geim}},\ }\bibfield  {title}
  {\bibinfo {title} {The electronic properties of graphene},\ }\href
  {https://doi.org/10.1103/RevModPhys.81.109} {\bibfield  {journal} {\bibinfo
  {journal} {Rev. Mod. Phys.}\ }\textbf {\bibinfo {volume} {81}},\ \bibinfo
  {pages} {109} (\bibinfo {year} {2009})}\BibitemShut {NoStop}%
\bibitem [{\citenamefont {Bradlyn}\ \emph {et~al.}(2016)\citenamefont
  {Bradlyn}, \citenamefont {Cano}, \citenamefont {Wang}, \citenamefont
  {Vergniory}, \citenamefont {Felser}, \citenamefont {Cava},\ and\
  \citenamefont {Bernevig}}]{Beyond-Dirac}%
  \BibitemOpen
  \bibfield  {author} {\bibinfo {author} {\bibfnamefont {B.}~\bibnamefont
  {Bradlyn}}, \bibinfo {author} {\bibfnamefont {J.}~\bibnamefont {Cano}},
  \bibinfo {author} {\bibfnamefont {Z.}~\bibnamefont {Wang}}, \bibinfo {author}
  {\bibfnamefont {M.~G.}\ \bibnamefont {Vergniory}}, \bibinfo {author}
  {\bibfnamefont {C.}~\bibnamefont {Felser}}, \bibinfo {author} {\bibfnamefont
  {R.~J.}\ \bibnamefont {Cava}},\ and\ \bibinfo {author} {\bibfnamefont
  {B.~A.}\ \bibnamefont {Bernevig}},\ }\bibfield  {title} {\bibinfo {title}
  {Beyond dirac and weyl fermions: Unconventional quasiparticles in
  conventional crystals},\ }\href {https://doi.org/10.1126/science.aaf5037}
  {\bibfield  {journal} {\bibinfo  {journal} {Science}\ }\textbf {\bibinfo
  {volume} {353}},\ \bibinfo {pages} {aaf5037} (\bibinfo {year} {2016})},\
  \Eprint
  {https://arxiv.org/abs/https://www.science.org/doi/pdf/10.1126/science.aaf5037}
  {https://www.science.org/doi/pdf/10.1126/science.aaf5037} \BibitemShut
  {NoStop}%
\bibitem [{\citenamefont {Leykam}\ \emph {et~al.}(2017)\citenamefont {Leykam},
  \citenamefont {Bliokh}, \citenamefont {Huang}, \citenamefont {Chong},\ and\
  \citenamefont {Nori}}]{non-H-topo}%
  \BibitemOpen
  \bibfield  {author} {\bibinfo {author} {\bibfnamefont {D.}~\bibnamefont
  {Leykam}}, \bibinfo {author} {\bibfnamefont {K.~Y.}\ \bibnamefont {Bliokh}},
  \bibinfo {author} {\bibfnamefont {C.}~\bibnamefont {Huang}}, \bibinfo
  {author} {\bibfnamefont {Y.~D.}\ \bibnamefont {Chong}},\ and\ \bibinfo
  {author} {\bibfnamefont {F.}~\bibnamefont {Nori}},\ }\bibfield  {title}
  {\bibinfo {title} {Edge modes, degeneracies, and topological numbers in
  non-hermitian systems},\ }\href
  {https://doi.org/10.1103/PhysRevLett.118.040401} {\bibfield  {journal}
  {\bibinfo  {journal} {Phys. Rev. Lett.}\ }\textbf {\bibinfo {volume} {118}},\
  \bibinfo {pages} {040401} (\bibinfo {year} {2017})}\BibitemShut {NoStop}%
\bibitem [{\citenamefont {Shen}\ \emph {et~al.}(2018)\citenamefont {Shen},
  \citenamefont {Zhen},\ and\ \citenamefont {Fu}}]{non-H-band-theory}%
  \BibitemOpen
  \bibfield  {author} {\bibinfo {author} {\bibfnamefont {H.}~\bibnamefont
  {Shen}}, \bibinfo {author} {\bibfnamefont {B.}~\bibnamefont {Zhen}},\ and\
  \bibinfo {author} {\bibfnamefont {L.}~\bibnamefont {Fu}},\ }\bibfield
  {title} {\bibinfo {title} {Topological band theory for non-hermitian
  hamiltonians},\ }\href {https://doi.org/10.1103/PhysRevLett.120.146402}
  {\bibfield  {journal} {\bibinfo  {journal} {Phys. Rev. Lett.}\ }\textbf
  {\bibinfo {volume} {120}},\ \bibinfo {pages} {146402} (\bibinfo {year}
  {2018})}\BibitemShut {NoStop}%
\bibitem [{\citenamefont {Yao}\ and\ \citenamefont
  {Wang}(2018)}]{non-H-edge-states}%
  \BibitemOpen
  \bibfield  {author} {\bibinfo {author} {\bibfnamefont {S.}~\bibnamefont
  {Yao}}\ and\ \bibinfo {author} {\bibfnamefont {Z.}~\bibnamefont {Wang}},\
  }\bibfield  {title} {\bibinfo {title} {Edge states and topological invariants
  of non-hermitian systems},\ }\href
  {https://doi.org/10.1103/PhysRevLett.121.086803} {\bibfield  {journal}
  {\bibinfo  {journal} {Phys. Rev. Lett.}\ }\textbf {\bibinfo {volume} {121}},\
  \bibinfo {pages} {086803} (\bibinfo {year} {2018})}\BibitemShut {NoStop}%
\bibitem [{\citenamefont {Xue}\ \emph {et~al.}(2020)\citenamefont {Xue},
  \citenamefont {Wang}, \citenamefont {Zhang},\ and\ \citenamefont
  {Chong}}]{non-H-DP}%
  \BibitemOpen
  \bibfield  {author} {\bibinfo {author} {\bibfnamefont {H.}~\bibnamefont
  {Xue}}, \bibinfo {author} {\bibfnamefont {Q.}~\bibnamefont {Wang}}, \bibinfo
  {author} {\bibfnamefont {B.}~\bibnamefont {Zhang}},\ and\ \bibinfo {author}
  {\bibfnamefont {Y.~D.}\ \bibnamefont {Chong}},\ }\bibfield  {title} {\bibinfo
  {title} {Non-hermitian dirac cones},\ }\href
  {https://doi.org/10.1103/PhysRevLett.124.236403} {\bibfield  {journal}
  {\bibinfo  {journal} {Phys. Rev. Lett.}\ }\textbf {\bibinfo {volume} {124}},\
  \bibinfo {pages} {236403} (\bibinfo {year} {2020})}\BibitemShut {NoStop}%
\bibitem [{\citenamefont {Wang}(2022)}]{special-EP}%
  \BibitemOpen
  \bibfield  {author} {\bibinfo {author} {\bibfnamefont {T.}~\bibnamefont
  {Wang}},\ }\bibfield  {title} {\bibinfo {title} {Special exceptional point
  acting as dirac point in one dimensional -symmetric photonic crystal},\
  }\href {https://doi.org/10.1088/1367-2630/ac9a9f} {\bibfield  {journal}
  {\bibinfo  {journal} {New Journal of Physics}\ }\textbf {\bibinfo {volume}
  {24}},\ \bibinfo {pages} {113016} (\bibinfo {year} {2022})}\BibitemShut
  {NoStop}%
\bibitem [{\citenamefont {Luo}\ \emph {et~al.}(2021)\citenamefont {Luo},
  \citenamefont {Shao}, \citenamefont {Li}, \citenamefont {Fan}, \citenamefont
  {Peng}, \citenamefont {Wang}, \citenamefont {Luo},\ and\ \citenamefont
  {Lai}}]{complex-DP}%
  \BibitemOpen
  \bibfield  {author} {\bibinfo {author} {\bibfnamefont {L.}~\bibnamefont
  {Luo}}, \bibinfo {author} {\bibfnamefont {Y.}~\bibnamefont {Shao}}, \bibinfo
  {author} {\bibfnamefont {J.}~\bibnamefont {Li}}, \bibinfo {author}
  {\bibfnamefont {R.}~\bibnamefont {Fan}}, \bibinfo {author} {\bibfnamefont
  {R.}~\bibnamefont {Peng}}, \bibinfo {author} {\bibfnamefont {M.}~\bibnamefont
  {Wang}}, \bibinfo {author} {\bibfnamefont {J.}~\bibnamefont {Luo}},\ and\
  \bibinfo {author} {\bibfnamefont {Y.}~\bibnamefont {Lai}},\ }\bibfield
  {title} {\bibinfo {title} {Non-hermitian effective medium theory and complex
  dirac-like cones},\ }\href {https://doi.org/10.1364/OE.425862} {\bibfield
  {journal} {\bibinfo  {journal} {Opt. Express}\ }\textbf {\bibinfo {volume}
  {29}},\ \bibinfo {pages} {14345} (\bibinfo {year} {2021})}\BibitemShut
  {NoStop}%
\bibitem [{\citenamefont {Zhen}\ \emph {et~al.}(2015)\citenamefont {Zhen},
  \citenamefont {Hsu}, \citenamefont {Igarashi}, \citenamefont {Lu},
  \citenamefont {Kaminer}, \citenamefont {Pick}, \citenamefont {Chua},
  \citenamefont {Joannopoulos},\ and\ \citenamefont {Soljačić}}]{rings-EPs}%
  \BibitemOpen
  \bibfield  {author} {\bibinfo {author} {\bibfnamefont {B.}~\bibnamefont
  {Zhen}}, \bibinfo {author} {\bibfnamefont {C.~W.}\ \bibnamefont {Hsu}},
  \bibinfo {author} {\bibfnamefont {Y.}~\bibnamefont {Igarashi}}, \bibinfo
  {author} {\bibfnamefont {L.}~\bibnamefont {Lu}}, \bibinfo {author}
  {\bibfnamefont {I.}~\bibnamefont {Kaminer}}, \bibinfo {author} {\bibfnamefont
  {A.}~\bibnamefont {Pick}}, \bibinfo {author} {\bibfnamefont {S.-L.}\
  \bibnamefont {Chua}}, \bibinfo {author} {\bibfnamefont {J.~D.}\ \bibnamefont
  {Joannopoulos}},\ and\ \bibinfo {author} {\bibfnamefont {M.}~\bibnamefont
  {Soljačić}},\ }\bibfield  {title} {\bibinfo {title} {Spawning rings of
  exceptional points out of dirac cones},\ }\href
  {https://doi.org/10.1038/nature14889} {\bibfield  {journal} {\bibinfo
  {journal} {Nature}\ }\textbf {\bibinfo {volume} {525}},\ \bibinfo {pages}
  {354} (\bibinfo {year} {2015})}\BibitemShut {NoStop}%
\bibitem [{\citenamefont {Özdemir}(2018)}]{ozdemir}%
  \BibitemOpen
  \bibfield  {author} {\bibinfo {author} {\bibfnamefont {S.~K.}\ \bibnamefont
  {Özdemir}},\ }\bibfield  {title} {\bibinfo {title} {Fermi arcs connect
  topological degeneracies},\ }\href {https://doi.org/10.1126/science.aar8210}
  {\bibfield  {journal} {\bibinfo  {journal} {Science}\ }\textbf {\bibinfo
  {volume} {359}},\ \bibinfo {pages} {995} (\bibinfo {year}
  {2018})}\BibitemShut {NoStop}%
\bibitem [{\citenamefont {Ashida}\ \emph {et~al.}(2020)\citenamefont {Ashida},
  \citenamefont {Gong},\ and\ \citenamefont {Ueda}}]{non-hermitian}%
  \BibitemOpen
  \bibfield  {author} {\bibinfo {author} {\bibfnamefont {Y.}~\bibnamefont
  {Ashida}}, \bibinfo {author} {\bibfnamefont {Z.}~\bibnamefont {Gong}},\ and\
  \bibinfo {author} {\bibfnamefont {M.}~\bibnamefont {Ueda}},\ }\bibfield
  {title} {\bibinfo {title} {Non-hermitian physics},\ }\href
  {https://doi.org/10.1080/00018732.2021.1876991} {\bibfield  {journal}
  {\bibinfo  {journal} {Advances in Physics}\ }\textbf {\bibinfo {volume}
  {69}},\ \bibinfo {pages} {249} (\bibinfo {year} {2020})}\BibitemShut
  {NoStop}%
\bibitem [{\citenamefont {Miri}\ and\ \citenamefont {Alù}(2019)}]{alu-review}%
  \BibitemOpen
  \bibfield  {author} {\bibinfo {author} {\bibfnamefont {M.-A.}\ \bibnamefont
  {Miri}}\ and\ \bibinfo {author} {\bibfnamefont {A.}~\bibnamefont {Alù}},\
  }\bibfield  {title} {\bibinfo {title} {Exceptional points in optics and
  photonics},\ }\href {https://doi.org/10.1126/science.aar7709} {\bibfield
  {journal} {\bibinfo  {journal} {Science}\ }\textbf {\bibinfo {volume}
  {363}},\ \bibinfo {pages} {eaar7709} (\bibinfo {year} {2019})}\BibitemShut
  {NoStop}%
\bibitem [{\citenamefont {Doppler}\ \emph {et~al.}(2016)\citenamefont
  {Doppler}, \citenamefont {Mailybaev}, \citenamefont {Böhm}, \citenamefont
  {Kuhl}, \citenamefont {Girschik}, \citenamefont {Libisch}, \citenamefont
  {Milburn}, \citenamefont {Rabl}, \citenamefont {Moiseyev},\ and\
  \citenamefont {Rotter}}]{Doppler}%
  \BibitemOpen
  \bibfield  {author} {\bibinfo {author} {\bibfnamefont {J.}~\bibnamefont
  {Doppler}}, \bibinfo {author} {\bibfnamefont {A.~A.}\ \bibnamefont
  {Mailybaev}}, \bibinfo {author} {\bibfnamefont {J.}~\bibnamefont {Böhm}},
  \bibinfo {author} {\bibfnamefont {U.}~\bibnamefont {Kuhl}}, \bibinfo {author}
  {\bibfnamefont {A.}~\bibnamefont {Girschik}}, \bibinfo {author}
  {\bibfnamefont {F.}~\bibnamefont {Libisch}}, \bibinfo {author} {\bibfnamefont
  {T.~J.}\ \bibnamefont {Milburn}}, \bibinfo {author} {\bibfnamefont
  {P.}~\bibnamefont {Rabl}}, \bibinfo {author} {\bibfnamefont {N.}~\bibnamefont
  {Moiseyev}},\ and\ \bibinfo {author} {\bibfnamefont {S.}~\bibnamefont
  {Rotter}},\ }\bibfield  {title} {\bibinfo {title} {Dynamically encircling an
  exceptional point for asymmetric mode switching},\ }\href
  {https://doi.org/10.1038/nature18605} {\bibfield  {journal} {\bibinfo
  {journal} {Nature}\ }\textbf {\bibinfo {volume} {537}},\ \bibinfo {pages}
  {76–79} (\bibinfo {year} {2016})}\BibitemShut {NoStop}%
\bibitem [{\citenamefont {Yoon}\ \emph {et~al.}(2018)\citenamefont {Yoon},
  \citenamefont {Choi}, \citenamefont {Hahn}, \citenamefont {Kim},
  \citenamefont {Song}, \citenamefont {Yang}, \citenamefont {Lee},
  \citenamefont {Kim}, \citenamefont {Lee}, \citenamefont {Shin}, \citenamefont
  {Lee},\ and\ \citenamefont {Berini}}]{yoon}%
  \BibitemOpen
  \bibfield  {author} {\bibinfo {author} {\bibfnamefont {J.~W.}\ \bibnamefont
  {Yoon}}, \bibinfo {author} {\bibfnamefont {Y.}~\bibnamefont {Choi}}, \bibinfo
  {author} {\bibfnamefont {C.}~\bibnamefont {Hahn}}, \bibinfo {author}
  {\bibfnamefont {G.}~\bibnamefont {Kim}}, \bibinfo {author} {\bibfnamefont
  {S.~H.}\ \bibnamefont {Song}}, \bibinfo {author} {\bibfnamefont {K.-Y.}\
  \bibnamefont {Yang}}, \bibinfo {author} {\bibfnamefont {J.~Y.}\ \bibnamefont
  {Lee}}, \bibinfo {author} {\bibfnamefont {Y.}~\bibnamefont {Kim}}, \bibinfo
  {author} {\bibfnamefont {C.~S.}\ \bibnamefont {Lee}}, \bibinfo {author}
  {\bibfnamefont {J.~K.}\ \bibnamefont {Shin}}, \bibinfo {author}
  {\bibfnamefont {H.-S.}\ \bibnamefont {Lee}},\ and\ \bibinfo {author}
  {\bibfnamefont {P.}~\bibnamefont {Berini}},\ }\bibfield  {title} {\bibinfo
  {title} {Time-asymmetric loop around an exceptional point over the full
  optical communications band},\ }\href
  {https://doi.org/10.1038/s41586-018-0523-2} {\bibfield  {journal} {\bibinfo
  {journal} {Nature}\ }\textbf {\bibinfo {volume} {562}},\ \bibinfo {pages}
  {86–90} (\bibinfo {year} {2018})}\BibitemShut {NoStop}%
\bibitem [{\citenamefont {Zhou}\ \emph {et~al.}(2018)\citenamefont {Zhou},
  \citenamefont {Peng}, \citenamefont {Yoon}, \citenamefont {Hsu},
  \citenamefont {Nelson}, \citenamefont {Fu}, \citenamefont {Joannopoulos},
  \citenamefont {Soljačić},\ and\ \citenamefont {Zhen}}]{Zhou-sci}%
  \BibitemOpen
  \bibfield  {author} {\bibinfo {author} {\bibfnamefont {H.}~\bibnamefont
  {Zhou}}, \bibinfo {author} {\bibfnamefont {C.}~\bibnamefont {Peng}}, \bibinfo
  {author} {\bibfnamefont {Y.}~\bibnamefont {Yoon}}, \bibinfo {author}
  {\bibfnamefont {C.~W.}\ \bibnamefont {Hsu}}, \bibinfo {author} {\bibfnamefont
  {K.~A.}\ \bibnamefont {Nelson}}, \bibinfo {author} {\bibfnamefont
  {L.}~\bibnamefont {Fu}}, \bibinfo {author} {\bibfnamefont {J.~D.}\
  \bibnamefont {Joannopoulos}}, \bibinfo {author} {\bibfnamefont
  {M.}~\bibnamefont {Soljačić}},\ and\ \bibinfo {author} {\bibfnamefont
  {B.}~\bibnamefont {Zhen}},\ }\bibfield  {title} {\bibinfo {title}
  {Observation of bulk fermi arc and polarization half charge from paired
  exceptional points},\ }\href {https://doi.org/10.1126/science.aap9859}
  {\bibfield  {journal} {\bibinfo  {journal} {Science}\ }\textbf {\bibinfo
  {volume} {359}},\ \bibinfo {pages} {1009} (\bibinfo {year}
  {2018})}\BibitemShut {NoStop}%
\bibitem [{\citenamefont {Hodaei}\ \emph {et~al.}(2017)\citenamefont {Hodaei},
  \citenamefont {Hassan}, \citenamefont {Wittek}, \citenamefont
  {Garcia-Gracia}, \citenamefont {El-Ganainy}, \citenamefont
  {Christodoulides},\ and\ \citenamefont {Khajavikhan}}]{hodaei}%
  \BibitemOpen
  \bibfield  {author} {\bibinfo {author} {\bibfnamefont {H.}~\bibnamefont
  {Hodaei}}, \bibinfo {author} {\bibfnamefont {A.~U.}\ \bibnamefont {Hassan}},
  \bibinfo {author} {\bibfnamefont {S.}~\bibnamefont {Wittek}}, \bibinfo
  {author} {\bibfnamefont {H.}~\bibnamefont {Garcia-Gracia}}, \bibinfo {author}
  {\bibfnamefont {R.}~\bibnamefont {El-Ganainy}}, \bibinfo {author}
  {\bibfnamefont {D.~N.}\ \bibnamefont {Christodoulides}},\ and\ \bibinfo
  {author} {\bibfnamefont {M.}~\bibnamefont {Khajavikhan}},\ }\bibfield
  {title} {\bibinfo {title} {Enhanced sensitivity at higher-order exceptional
  points},\ }\href {https://doi.org/10.1038/nature23280} {\bibfield  {journal}
  {\bibinfo  {journal} {Nature}\ }\textbf {\bibinfo {volume} {548}},\ \bibinfo
  {pages} {187} (\bibinfo {year} {2017})}\BibitemShut {NoStop}%
\bibitem [{\citenamefont {Chen}\ \emph {et~al.}(2017)\citenamefont {Chen},
  \citenamefont {Özdemir}, \citenamefont {Zhao}, \citenamefont {Wiersig},\
  and\ \citenamefont {Yang}}]{chen-nat17}%
  \BibitemOpen
  \bibfield  {author} {\bibinfo {author} {\bibfnamefont {W.}~\bibnamefont
  {Chen}}, \bibinfo {author} {\bibfnamefont {S.~K.}\ \bibnamefont {Özdemir}},
  \bibinfo {author} {\bibfnamefont {G.}~\bibnamefont {Zhao}}, \bibinfo {author}
  {\bibfnamefont {J.}~\bibnamefont {Wiersig}},\ and\ \bibinfo {author}
  {\bibfnamefont {L.}~\bibnamefont {Yang}},\ }\bibfield  {title} {\bibinfo
  {title} {Exceptional points enhance sensing in an optical microcavity},\
  }\href {https://doi.org/10.1038/nature23281} {\bibfield  {journal} {\bibinfo
  {journal} {Nature}\ }\textbf {\bibinfo {volume} {548}},\ \bibinfo {pages}
  {192} (\bibinfo {year} {2017})}\BibitemShut {NoStop}%
\bibitem [{\citenamefont {Peng}\ \emph {et~al.}(2016)\citenamefont {Peng},
  \citenamefont {Özdemir}, \citenamefont {Liertzer}, \citenamefont {Chen},
  \citenamefont {Kramer}, \citenamefont {Yılmaz}, \citenamefont {Wiersig},
  \citenamefont {Rotter},\ and\ \citenamefont {Yang}}]{pengPNAS}%
  \BibitemOpen
  \bibfield  {author} {\bibinfo {author} {\bibfnamefont {B.}~\bibnamefont
  {Peng}}, \bibinfo {author} {\bibfnamefont {S.~K.}\ \bibnamefont {Özdemir}},
  \bibinfo {author} {\bibfnamefont {M.}~\bibnamefont {Liertzer}}, \bibinfo
  {author} {\bibfnamefont {W.}~\bibnamefont {Chen}}, \bibinfo {author}
  {\bibfnamefont {J.}~\bibnamefont {Kramer}}, \bibinfo {author} {\bibfnamefont
  {H.}~\bibnamefont {Yılmaz}}, \bibinfo {author} {\bibfnamefont
  {J.}~\bibnamefont {Wiersig}}, \bibinfo {author} {\bibfnamefont
  {S.}~\bibnamefont {Rotter}},\ and\ \bibinfo {author} {\bibfnamefont
  {L.}~\bibnamefont {Yang}},\ }\bibfield  {title} {\bibinfo {title} {Chiral
  modes and directional lasing at exceptional points},\ }\href
  {https://doi.org/10.1073/pnas.1603318113} {\bibfield  {journal} {\bibinfo
  {journal} {Proceedings of the National Academy of Sciences}\ }\textbf
  {\bibinfo {volume} {113}},\ \bibinfo {pages} {6845} (\bibinfo {year}
  {2016})}\BibitemShut {NoStop}%
\bibitem [{\citenamefont {von Neuman}\ and\ \citenamefont
  {Wigner}(1929)}]{Neuman1929}%
  \BibitemOpen
  \bibfield  {author} {\bibinfo {author} {\bibfnamefont {J.}~\bibnamefont {von
  Neuman}}\ and\ \bibinfo {author} {\bibfnamefont {E.}~\bibnamefont {Wigner}},\
  }\bibfield  {title} {\bibinfo {title} {Uber merkwürdige diskrete
  {Eigenwerte}. {Uber} das {Verhalten} von {Eigenwerten} bei adiabatischen
  {Prozessen}},\ }\href@noop {} {\bibfield  {journal} {\bibinfo  {journal}
  {Zeitschrift für Physik}\ }\textbf {\bibinfo {volume} {30}},\ \bibinfo
  {pages} {467} (\bibinfo {year} {1929})}\BibitemShut {NoStop}%
\bibitem [{\citenamefont {Stillinger}\ and\ \citenamefont
  {Herrick}(1975)}]{Stillinger1975}%
  \BibitemOpen
  \bibfield  {author} {\bibinfo {author} {\bibfnamefont {F.~H.}\ \bibnamefont
  {Stillinger}}\ and\ \bibinfo {author} {\bibfnamefont {D.~R.}\ \bibnamefont
  {Herrick}},\ }\bibfield  {title} {\bibinfo {title} {Bound states in the
  continuum},\ }\href {https://link.aps.org/doi/10.1103/PhysRevA.11.446}
  {\bibfield  {journal} {\bibinfo  {journal} {Physical Review A}\ }\textbf
  {\bibinfo {volume} {11}},\ \bibinfo {pages} {446} (\bibinfo {year}
  {1975})}\BibitemShut {NoStop}%
\bibitem [{\citenamefont {Hsu}\ \emph {et~al.}(2016)\citenamefont {Hsu},
  \citenamefont {Zhen}, \citenamefont {Stone}, \citenamefont {Joannopoulos},\
  and\ \citenamefont {Soljačić}}]{Hsu2016}%
  \BibitemOpen
  \bibfield  {author} {\bibinfo {author} {\bibfnamefont {C.~W.}\ \bibnamefont
  {Hsu}}, \bibinfo {author} {\bibfnamefont {B.}~\bibnamefont {Zhen}}, \bibinfo
  {author} {\bibfnamefont {A.~D.}\ \bibnamefont {Stone}}, \bibinfo {author}
  {\bibfnamefont {J.~D.}\ \bibnamefont {Joannopoulos}},\ and\ \bibinfo {author}
  {\bibfnamefont {M.}~\bibnamefont {Soljačić}},\ }\bibfield  {title}
  {\bibinfo {title} {Bound states in the continuum},\ }\bibfield  {journal}
  {\bibinfo  {journal} {Nature Reviews Materials}\ }\textbf {\bibinfo {volume}
  {1}},\ \href {https://doi.org/10.1038/natrevmats.2016.4}
  {10.1038/natrevmats.2016.4} (\bibinfo {year} {2016})\BibitemShut {NoStop}%
\bibitem [{\citenamefont {Parker}(1966)}]{Parker1966}%
  \BibitemOpen
  \bibfield  {author} {\bibinfo {author} {\bibfnamefont {R.}~\bibnamefont
  {Parker}},\ }\bibfield  {title} {\bibinfo {title} {Resonance effects in wake
  shedding from parallel plates: {Some} experimental observations},\ }\href
  {http://www.sciencedirect.com/science/article/pii/0022460X66901544}
  {\bibfield  {journal} {\bibinfo  {journal} {J. Sound Vib.}\ }\textbf
  {\bibinfo {volume} {4}},\ \bibinfo {pages} {62 } (\bibinfo {year}
  {1966})}\BibitemShut {NoStop}%
\bibitem [{\citenamefont {Kim}\ \emph {et~al.}(1999)\citenamefont {Kim},
  \citenamefont {Satanin}, \citenamefont {Joe},\ and\ \citenamefont
  {Cosby}}]{Kim1999}%
  \BibitemOpen
  \bibfield  {author} {\bibinfo {author} {\bibfnamefont {C.~S.}\ \bibnamefont
  {Kim}}, \bibinfo {author} {\bibfnamefont {A.~M.}\ \bibnamefont {Satanin}},
  \bibinfo {author} {\bibfnamefont {Y.~S.}\ \bibnamefont {Joe}},\ and\ \bibinfo
  {author} {\bibfnamefont {R.~M.}\ \bibnamefont {Cosby}},\ }\bibfield  {title}
  {\bibinfo {title} {Resonant tunneling in a quantum waveguide: Effect of a
  finite-size attractive impurity},\ }\href
  {https://doi.org/10.1103/PhysRevB.60.10962} {\bibfield  {journal} {\bibinfo
  {journal} {Phys. Rev. B}\ }\textbf {\bibinfo {volume} {60}},\ \bibinfo
  {pages} {10962} (\bibinfo {year} {1999})}\BibitemShut {NoStop}%
\bibitem [{\citenamefont {Marinica}\ \emph {et~al.}(2008)\citenamefont
  {Marinica}, \citenamefont {Borisov},\ and\ \citenamefont
  {Shabanov}}]{Marinica2008}%
  \BibitemOpen
  \bibfield  {author} {\bibinfo {author} {\bibfnamefont {D.~C.}\ \bibnamefont
  {Marinica}}, \bibinfo {author} {\bibfnamefont {A.~G.}\ \bibnamefont
  {Borisov}},\ and\ \bibinfo {author} {\bibfnamefont {S.~V.}\ \bibnamefont
  {Shabanov}},\ }\bibfield  {title} {\bibinfo {title} {Bound states in the
  continuum in photonics},\ }\href
  {https://doi.org/10.1103/PhysRevLett.100.183902} {\bibfield  {journal}
  {\bibinfo  {journal} {Physical Review Letters}\ }\textbf {\bibinfo {volume}
  {100}},\ \bibinfo {pages} {183902} (\bibinfo {year} {2008})}\BibitemShut
  {NoStop}%
\bibitem [{\citenamefont {Bulgakov}\ and\ \citenamefont
  {Sadreev}(2008)}]{Bulgakov2008}%
  \BibitemOpen
  \bibfield  {author} {\bibinfo {author} {\bibfnamefont {E.~N.}\ \bibnamefont
  {Bulgakov}}\ and\ \bibinfo {author} {\bibfnamefont {A.~F.}\ \bibnamefont
  {Sadreev}},\ }\bibfield  {title} {\bibinfo {title} {Bound states in the
  continuum in photonic waveguides inspired by defects},\ }\href
  {https://doi.org/10.1103/PhysRevB.78.075105} {\bibfield  {journal} {\bibinfo
  {journal} {Physical Review B}\ }\textbf {\bibinfo {volume} {78}},\ \bibinfo
  {pages} {075105} (\bibinfo {year} {2008})}\BibitemShut {NoStop}%
\bibitem [{\citenamefont {Plotnik}\ \emph {et~al.}(2011)\citenamefont
  {Plotnik}, \citenamefont {Peleg}, \citenamefont {Dreisow}, \citenamefont
  {Heinrich}, \citenamefont {Nolte}, \citenamefont {Szameit},\ and\
  \citenamefont {Segev}}]{Plotnik2011}%
  \BibitemOpen
  \bibfield  {author} {\bibinfo {author} {\bibfnamefont {Y.}~\bibnamefont
  {Plotnik}}, \bibinfo {author} {\bibfnamefont {O.}~\bibnamefont {Peleg}},
  \bibinfo {author} {\bibfnamefont {F.}~\bibnamefont {Dreisow}}, \bibinfo
  {author} {\bibfnamefont {M.}~\bibnamefont {Heinrich}}, \bibinfo {author}
  {\bibfnamefont {S.}~\bibnamefont {Nolte}}, \bibinfo {author} {\bibfnamefont
  {A.}~\bibnamefont {Szameit}},\ and\ \bibinfo {author} {\bibfnamefont
  {M.}~\bibnamefont {Segev}},\ }\bibfield  {title} {\bibinfo {title}
  {Experimental observation of optical bound states in the continuum},\ }\href
  {https://doi.org/10.1103/PhysRevLett.107.183901} {\bibfield  {journal}
  {\bibinfo  {journal} {Physical Review Letters}\ }\textbf {\bibinfo {volume}
  {107}},\ \bibinfo {pages} {183901} (\bibinfo {year} {2011})}\BibitemShut
  {NoStop}%
\bibitem [{\citenamefont {Hsu}\ \emph {et~al.}(2013)\citenamefont {Hsu},
  \citenamefont {Zhen}, \citenamefont {Lee}, \citenamefont {Chua},
  \citenamefont {Johnson}, \citenamefont {Joannopoulos},\ and\ \citenamefont
  {Soljačić}}]{Hsu2013}%
  \BibitemOpen
  \bibfield  {author} {\bibinfo {author} {\bibfnamefont {C.~W.}\ \bibnamefont
  {Hsu}}, \bibinfo {author} {\bibfnamefont {B.}~\bibnamefont {Zhen}}, \bibinfo
  {author} {\bibfnamefont {J.}~\bibnamefont {Lee}}, \bibinfo {author}
  {\bibfnamefont {S.-L.}\ \bibnamefont {Chua}}, \bibinfo {author}
  {\bibfnamefont {S.~G.}\ \bibnamefont {Johnson}}, \bibinfo {author}
  {\bibfnamefont {J.~D.}\ \bibnamefont {Joannopoulos}},\ and\ \bibinfo {author}
  {\bibfnamefont {M.}~\bibnamefont {Soljačić}},\ }\bibfield  {title}
  {\bibinfo {title} {Observation of trapped light within the radiation
  continuum},\ }\href {https://doi.org/10.1038/nature12289} {\bibfield
  {journal} {\bibinfo  {journal} {Nature}\ }\textbf {\bibinfo {volume} {499}},\
  \bibinfo {pages} {188} (\bibinfo {year} {2013})}\BibitemShut {NoStop}%
\bibitem [{\citenamefont {Monticone}\ and\ \citenamefont
  {Al\`u}(2014)}]{Monticone2014}%
  \BibitemOpen
  \bibfield  {author} {\bibinfo {author} {\bibfnamefont {F.}~\bibnamefont
  {Monticone}}\ and\ \bibinfo {author} {\bibfnamefont {A.}~\bibnamefont
  {Al\`u}},\ }\bibfield  {title} {\bibinfo {title} {Embedded photonic
  eigenvalues in 3d nanostructures},\ }\href
  {https://doi.org/10.1103/PhysRevLett.112.213903} {\bibfield  {journal}
  {\bibinfo  {journal} {Physical Review Letters}\ }\textbf {\bibinfo {volume}
  {112}},\ \bibinfo {pages} {213903} (\bibinfo {year} {2014})}\BibitemShut
  {NoStop}%
\bibitem [{\citenamefont {Kikkawa}\ \emph {et~al.}(2020)\citenamefont
  {Kikkawa}, \citenamefont {Nishida},\ and\ \citenamefont
  {Kadoya}}]{Kikkawa_2020}%
  \BibitemOpen
  \bibfield  {author} {\bibinfo {author} {\bibfnamefont {R.}~\bibnamefont
  {Kikkawa}}, \bibinfo {author} {\bibfnamefont {M.}~\bibnamefont {Nishida}},\
  and\ \bibinfo {author} {\bibfnamefont {Y.}~\bibnamefont {Kadoya}},\
  }\bibfield  {title} {\bibinfo {title} {Bound states in the continuum and
  exceptional points in dielectric waveguide equipped with a metal grating},\
  }\href {https://doi.org/10.1088/1367-2630/ab97e9} {\bibfield  {journal}
  {\bibinfo  {journal} {New Journal of Physics}\ }\textbf {\bibinfo {volume}
  {22}},\ \bibinfo {pages} {073029} (\bibinfo {year} {2020})}\BibitemShut
  {NoStop}%
\bibitem [{\citenamefont {Cerjan}\ \emph {et~al.}(2019)\citenamefont {Cerjan},
  \citenamefont {Hsu},\ and\ \citenamefont {Rechtsman}}]{Cerjan2021}%
  \BibitemOpen
  \bibfield  {author} {\bibinfo {author} {\bibfnamefont {A.}~\bibnamefont
  {Cerjan}}, \bibinfo {author} {\bibfnamefont {C.~W.}\ \bibnamefont {Hsu}},\
  and\ \bibinfo {author} {\bibfnamefont {M.~C.}\ \bibnamefont {Rechtsman}},\
  }\bibfield  {title} {\bibinfo {title} {Bound states in the continuum through
  environmental design},\ }\href
  {https://doi.org/10.1103/PhysRevLett.123.023902} {\bibfield  {journal}
  {\bibinfo  {journal} {Phys. Rev. Lett.}\ }\textbf {\bibinfo {volume} {123}},\
  \bibinfo {pages} {023902} (\bibinfo {year} {2019})}\BibitemShut {NoStop}%
\bibitem [{\citenamefont {Qin}\ \emph {et~al.}(2022)\citenamefont {Qin},
  \citenamefont {Shi},\ and\ \citenamefont {Ou}}]{shi2022}%
  \BibitemOpen
  \bibfield  {author} {\bibinfo {author} {\bibfnamefont {H.}~\bibnamefont
  {Qin}}, \bibinfo {author} {\bibfnamefont {X.}~\bibnamefont {Shi}},\ and\
  \bibinfo {author} {\bibfnamefont {H.}~\bibnamefont {Ou}},\ }\bibfield
  {title} {\bibinfo {title} {Exceptional points at bound states in the
  continuum in photonic integrated circuits},\ }\bibfield  {journal} {\bibinfo
  {journal} {Nanophotonics}\ }\href
  {https://doi.org/doi:10.1515/nanoph-2022-0420} {doi:10.1515/nanoph-2022-0420}
  (\bibinfo {year} {2022})\BibitemShut {NoStop}%
\bibitem [{\citenamefont {Kane}\ and\ \citenamefont {Mele}(2005)}]{Qhallgraph}%
  \BibitemOpen
  \bibfield  {author} {\bibinfo {author} {\bibfnamefont {C.~L.}\ \bibnamefont
  {Kane}}\ and\ \bibinfo {author} {\bibfnamefont {E.~J.}\ \bibnamefont
  {Mele}},\ }\bibfield  {title} {\bibinfo {title} {Quantum spin hall effect in
  graphene},\ }\href {https://doi.org/10.1103/PhysRevLett.95.226801} {\bibfield
   {journal} {\bibinfo  {journal} {Phys. Rev. Lett.}\ }\textbf {\bibinfo
  {volume} {95}},\ \bibinfo {pages} {226801} (\bibinfo {year}
  {2005})}\BibitemShut {NoStop}%
\bibitem [{\citenamefont {Haldane}\ and\ \citenamefont
  {Raghu}(2008)}]{edgephoton}%
  \BibitemOpen
  \bibfield  {author} {\bibinfo {author} {\bibfnamefont {F.~D.~M.}\
  \bibnamefont {Haldane}}\ and\ \bibinfo {author} {\bibfnamefont
  {S.}~\bibnamefont {Raghu}},\ }\bibfield  {title} {\bibinfo {title} {Possible
  realization of directional optical waveguides in photonic crystals with
  broken time-reversal symmetry},\ }\href
  {https://doi.org/10.1103/PhysRevLett.100.013904} {\bibfield  {journal}
  {\bibinfo  {journal} {Phys. Rev. Lett.}\ }\textbf {\bibinfo {volume} {100}},\
  \bibinfo {pages} {013904} (\bibinfo {year} {2008})}\BibitemShut {NoStop}%
\bibitem [{\citenamefont {Ozawa}\ \emph {et~al.}(2019)\citenamefont {Ozawa},
  \citenamefont {Price}, \citenamefont {Amo}, \citenamefont {Goldman},
  \citenamefont {Hafezi}, \citenamefont {Lu}, \citenamefont {Rechtsman},
  \citenamefont {Schuster}, \citenamefont {Simon}, \citenamefont {Zilberberg},\
  and\ \citenamefont {Carusotto}}]{RevModPhys}%
  \BibitemOpen
  \bibfield  {author} {\bibinfo {author} {\bibfnamefont {T.}~\bibnamefont
  {Ozawa}}, \bibinfo {author} {\bibfnamefont {H.~M.}\ \bibnamefont {Price}},
  \bibinfo {author} {\bibfnamefont {A.}~\bibnamefont {Amo}}, \bibinfo {author}
  {\bibfnamefont {N.}~\bibnamefont {Goldman}}, \bibinfo {author} {\bibfnamefont
  {M.}~\bibnamefont {Hafezi}}, \bibinfo {author} {\bibfnamefont
  {L.}~\bibnamefont {Lu}}, \bibinfo {author} {\bibfnamefont {M.~C.}\
  \bibnamefont {Rechtsman}}, \bibinfo {author} {\bibfnamefont {D.}~\bibnamefont
  {Schuster}}, \bibinfo {author} {\bibfnamefont {J.}~\bibnamefont {Simon}},
  \bibinfo {author} {\bibfnamefont {O.}~\bibnamefont {Zilberberg}},\ and\
  \bibinfo {author} {\bibfnamefont {I.}~\bibnamefont {Carusotto}},\ }\bibfield
  {title} {\bibinfo {title} {Topological photonics},\ }\href
  {https://doi.org/10.1103/RevModPhys.91.015006} {\bibfield  {journal}
  {\bibinfo  {journal} {Rev. Mod. Phys.}\ }\textbf {\bibinfo {volume} {91}},\
  \bibinfo {pages} {015006} (\bibinfo {year} {2019})}\BibitemShut {NoStop}%
\bibitem [{\citenamefont {Jin}\ \emph {et~al.}(2019)\citenamefont {Jin},
  \citenamefont {Yin}, \citenamefont {Ni}, \citenamefont {Soljačić},
  \citenamefont {Zhen},\ and\ \citenamefont {Peng}}]{mergingBICs}%
  \BibitemOpen
  \bibfield  {author} {\bibinfo {author} {\bibfnamefont {J.}~\bibnamefont
  {Jin}}, \bibinfo {author} {\bibfnamefont {X.}~\bibnamefont {Yin}}, \bibinfo
  {author} {\bibfnamefont {L.}~\bibnamefont {Ni}}, \bibinfo {author}
  {\bibfnamefont {M.}~\bibnamefont {Soljačić}}, \bibinfo {author}
  {\bibfnamefont {B.}~\bibnamefont {Zhen}},\ and\ \bibinfo {author}
  {\bibfnamefont {C.}~\bibnamefont {Peng}},\ }\bibfield  {title} {\bibinfo
  {title} {Topologically enabled ultrahigh-q guided resonances robust to
  out-of-plane scattering},\ }\href {https://doi.org/10.1038/s41586-019-1664-7}
  {\bibfield  {journal} {\bibinfo  {journal} {Nature}\ }\textbf {\bibinfo
  {volume} {574}},\ \bibinfo {pages} {501 } (\bibinfo {year}
  {2019})}\BibitemShut {NoStop}%
\bibitem [{\citenamefont {Smith}\ and\ \citenamefont
  {Schurig}(2003)}]{hypPRL-03}%
  \BibitemOpen
  \bibfield  {author} {\bibinfo {author} {\bibfnamefont {D.~R.}\ \bibnamefont
  {Smith}}\ and\ \bibinfo {author} {\bibfnamefont {D.}~\bibnamefont
  {Schurig}},\ }\bibfield  {title} {\bibinfo {title} {Electromagnetic wave
  propagation in media with indefinite permittivity and permeability tensors},\
  }\href {https://doi.org/10.1103/PhysRevLett.90.077405} {\bibfield  {journal}
  {\bibinfo  {journal} {Phys. Rev. Lett.}\ }\textbf {\bibinfo {volume} {90}},\
  \bibinfo {pages} {077405} (\bibinfo {year} {2003})}\BibitemShut {NoStop}%
\bibitem [{\citenamefont {Yao}\ \emph {et~al.}(2008)\citenamefont {Yao},
  \citenamefont {Liu}, \citenamefont {Liu}, \citenamefont {Wang}, \citenamefont
  {Sun}, \citenamefont {Bartal}, \citenamefont {Stacy},\ and\ \citenamefont
  {Zhang}}]{nanowiresSC08}%
  \BibitemOpen
  \bibfield  {author} {\bibinfo {author} {\bibfnamefont {J.}~\bibnamefont
  {Yao}}, \bibinfo {author} {\bibfnamefont {Z.}~\bibnamefont {Liu}}, \bibinfo
  {author} {\bibfnamefont {Y.}~\bibnamefont {Liu}}, \bibinfo {author}
  {\bibfnamefont {Y.}~\bibnamefont {Wang}}, \bibinfo {author} {\bibfnamefont
  {C.}~\bibnamefont {Sun}}, \bibinfo {author} {\bibfnamefont {G.}~\bibnamefont
  {Bartal}}, \bibinfo {author} {\bibfnamefont {A.~M.}\ \bibnamefont {Stacy}},\
  and\ \bibinfo {author} {\bibfnamefont {X.}~\bibnamefont {Zhang}},\ }\bibfield
   {title} {\bibinfo {title} {Optical negative refraction in bulk metamaterials
  of nanowires},\ }\href {https://doi.org/10.1126/science.1157566} {\bibfield
  {journal} {\bibinfo  {journal} {Science}\ }\textbf {\bibinfo {volume}
  {321}},\ \bibinfo {pages} {930} (\bibinfo {year} {2008})}\BibitemShut
  {NoStop}%
\bibitem [{\citenamefont {Noginov}\ \emph {et~al.}(2009)\citenamefont
  {Noginov}, \citenamefont {Barnakov}, \citenamefont {Zhu}, \citenamefont
  {Tumkur}, \citenamefont {Li},\ and\ \citenamefont {Narimanov}}]{bulkhyp}%
  \BibitemOpen
  \bibfield  {author} {\bibinfo {author} {\bibfnamefont {M.~A.}\ \bibnamefont
  {Noginov}}, \bibinfo {author} {\bibfnamefont {Y.~A.}\ \bibnamefont
  {Barnakov}}, \bibinfo {author} {\bibfnamefont {G.}~\bibnamefont {Zhu}},
  \bibinfo {author} {\bibfnamefont {T.}~\bibnamefont {Tumkur}}, \bibinfo
  {author} {\bibfnamefont {H.}~\bibnamefont {Li}},\ and\ \bibinfo {author}
  {\bibfnamefont {E.~E.}\ \bibnamefont {Narimanov}},\ }\bibfield  {title}
  {\bibinfo {title} {Bulk photonic metamaterial with hyperbolic dispersion},\
  }\href {https://doi.org/10.1063/1.3115145} {\bibfield  {journal} {\bibinfo
  {journal} {Applied Physics Letters}\ }\textbf {\bibinfo {volume} {94}},\
  \bibinfo {pages} {151105} (\bibinfo {year} {2009})}\BibitemShut {NoStop}%
\bibitem [{\citenamefont {Takayama}\ and\ \citenamefont
  {Lavrinenko}(2019)}]{Takayama:19}%
  \BibitemOpen
  \bibfield  {author} {\bibinfo {author} {\bibfnamefont {O.}~\bibnamefont
  {Takayama}}\ and\ \bibinfo {author} {\bibfnamefont {A.~V.}\ \bibnamefont
  {Lavrinenko}},\ }\bibfield  {title} {\bibinfo {title} {Optics with hyperbolic
  materials},\ }\href {https://doi.org/10.1364/JOSAB.36.000F38} {\bibfield
  {journal} {\bibinfo  {journal} {J. Opt. Soc. Am. B}\ }\textbf {\bibinfo
  {volume} {36}},\ \bibinfo {pages} {F38} (\bibinfo {year} {2019})}\BibitemShut
  {NoStop}%
\bibitem [{\citenamefont {Gomis-Bresco}\ \emph {et~al.}(2017)\citenamefont
  {Gomis-Bresco}, \citenamefont {Artigas},\ and\ \citenamefont
  {Torner}}]{Gomis-Bresco2017}%
  \BibitemOpen
  \bibfield  {author} {\bibinfo {author} {\bibfnamefont {J.}~\bibnamefont
  {Gomis-Bresco}}, \bibinfo {author} {\bibfnamefont {D.}~\bibnamefont
  {Artigas}},\ and\ \bibinfo {author} {\bibfnamefont {L.}~\bibnamefont
  {Torner}},\ }\bibfield  {title} {\bibinfo {title} {Anisotropy-induced
  photonic bound states in the continuum},\ }\href
  {https://doi.org/10.1038/nphoton.2017.31} {\bibfield  {journal} {\bibinfo
  {journal} {Nature Photonics}\ }\textbf {\bibinfo {volume} {11}},\ \bibinfo
  {pages} {232} (\bibinfo {year} {2017})}\BibitemShut {NoStop}%
\bibitem [{\citenamefont {Gomis-Bresco}\ \emph {et~al.}(2019)\citenamefont
  {Gomis-Bresco}, \citenamefont {Artigas},\ and\ \citenamefont
  {Torner}}]{GomisDPEP}%
  \BibitemOpen
  \bibfield  {author} {\bibinfo {author} {\bibfnamefont {J.}~\bibnamefont
  {Gomis-Bresco}}, \bibinfo {author} {\bibfnamefont {D.}~\bibnamefont
  {Artigas}},\ and\ \bibinfo {author} {\bibfnamefont {L.}~\bibnamefont
  {Torner}},\ }\bibfield  {title} {\bibinfo {title} {Transition from dirac
  points to exceptional points in anisotropic waveguides},\ }\href
  {https://doi.org/10.1103/PhysRevResearch.1.033010} {\bibfield  {journal}
  {\bibinfo  {journal} {Phys. Rev. Research}\ }\textbf {\bibinfo {volume}
  {1}},\ \bibinfo {pages} {033010} (\bibinfo {year} {2019})}\BibitemShut
  {NoStop}%
\bibitem [{\citenamefont {Mukherjee}\ \emph {et~al.}(2018)\citenamefont
  {Mukherjee}, \citenamefont {Gomis-Bresco}, \citenamefont {Pujol-Closa},
  \citenamefont {Artigas},\ and\ \citenamefont {Torner}}]{Mukherjee2018}%
  \BibitemOpen
  \bibfield  {author} {\bibinfo {author} {\bibfnamefont {S.}~\bibnamefont
  {Mukherjee}}, \bibinfo {author} {\bibfnamefont {J.}~\bibnamefont
  {Gomis-Bresco}}, \bibinfo {author} {\bibfnamefont {P.}~\bibnamefont
  {Pujol-Closa}}, \bibinfo {author} {\bibfnamefont {D.}~\bibnamefont
  {Artigas}},\ and\ \bibinfo {author} {\bibfnamefont {L.}~\bibnamefont
  {Torner}},\ }\bibfield  {title} {\bibinfo {title} {Topological properties of
  bound states in the continuum in geometries with broken anisotropy
  symmetry},\ }\href {https://doi.org/10.1103/PhysRevA.98.063826} {\bibfield
  {journal} {\bibinfo  {journal} {Physical Review A}\ }\textbf {\bibinfo
  {volume} {98}},\ \bibinfo {pages} {063826} (\bibinfo {year}
  {2018})}\BibitemShut {NoStop}%
\bibitem [{\citenamefont {Mukherjee}\ \emph {et~al.}(2022)\citenamefont
  {Mukherjee}, \citenamefont {Artigas},\ and\ \citenamefont
  {Torner}}]{Mukherjee-dya}%
  \BibitemOpen
  \bibfield  {author} {\bibinfo {author} {\bibfnamefont {S.}~\bibnamefont
  {Mukherjee}}, \bibinfo {author} {\bibfnamefont {D.}~\bibnamefont {Artigas}},\
  and\ \bibinfo {author} {\bibfnamefont {L.}~\bibnamefont {Torner}},\
  }\bibfield  {title} {\bibinfo {title} {Surface bound states in the continuum
  in dyakonov structures},\ }\href
  {https://doi.org/10.1103/PhysRevB.105.L201406} {\bibfield  {journal}
  {\bibinfo  {journal} {Phys. Rev. B}\ }\textbf {\bibinfo {volume} {105}},\
  \bibinfo {pages} {L201406} (\bibinfo {year} {2022})}\BibitemShut {NoStop}%
\bibitem [{\citenamefont {Pujol-Closa}\ \emph {et~al.}(2021)\citenamefont
  {Pujol-Closa}, \citenamefont {Gomis-Bresco}, \citenamefont {Mukherjee},
  \citenamefont {G\'{o}mez-D\'{i}az}, \citenamefont {Torner},\ and\
  \citenamefont {Artigas}}]{Pujol-Closa2021}%
  \BibitemOpen
  \bibfield  {author} {\bibinfo {author} {\bibfnamefont {P.}~\bibnamefont
  {Pujol-Closa}}, \bibinfo {author} {\bibfnamefont {J.}~\bibnamefont
  {Gomis-Bresco}}, \bibinfo {author} {\bibfnamefont {S.}~\bibnamefont
  {Mukherjee}}, \bibinfo {author} {\bibfnamefont {J.~S.}\ \bibnamefont
  {G\'{o}mez-D\'{i}az}}, \bibinfo {author} {\bibfnamefont {L.}~\bibnamefont
  {Torner}},\ and\ \bibinfo {author} {\bibfnamefont {D.}~\bibnamefont
  {Artigas}},\ }\bibfield  {title} {\bibinfo {title} {Slow light mediated by
  mode topological transitions in hyperbolic waveguides},\ }\href
  {https://doi.org/10.1364/OL.410423} {\bibfield  {journal} {\bibinfo
  {journal} {Optics Letters}\ }\textbf {\bibinfo {volume} {46}},\ \bibinfo
  {pages} {58} (\bibinfo {year} {2021})}\BibitemShut {NoStop}%
\bibitem [{\citenamefont {Mukherjee}\ \emph {et~al.}(2019)\citenamefont
  {Mukherjee}, \citenamefont {Gomis-Bresco}, \citenamefont {Pujol-Closa},
  \citenamefont {Artigas},\ and\ \citenamefont {Torner}}]{Mukherjee2019}%
  \BibitemOpen
  \bibfield  {author} {\bibinfo {author} {\bibfnamefont {S.}~\bibnamefont
  {Mukherjee}}, \bibinfo {author} {\bibfnamefont {J.}~\bibnamefont
  {Gomis-Bresco}}, \bibinfo {author} {\bibfnamefont {P.}~\bibnamefont
  {Pujol-Closa}}, \bibinfo {author} {\bibfnamefont {D.}~\bibnamefont
  {Artigas}},\ and\ \bibinfo {author} {\bibfnamefont {L.}~\bibnamefont
  {Torner}},\ }\bibfield  {title} {\bibinfo {title} {Angular control of
  anisotropy-induced bound states in the continuum},\ }\href
  {https://doi.org/10.1364/OL.44.005362} {\bibfield  {journal} {\bibinfo
  {journal} {Optics Letters}\ }\textbf {\bibinfo {volume} {44}},\ \bibinfo
  {pages} {5362} (\bibinfo {year} {2019})}\BibitemShut {NoStop}%
\bibitem [{\citenamefont {Mahmoodi}\ \emph {et~al.}(2019)\citenamefont
  {Mahmoodi}, \citenamefont {Tavassoli}, \citenamefont {Takayama},
  \citenamefont {Sukham}, \citenamefont {Malureanu},\ and\ \citenamefont
  {Lavrinenko}}]{Lavrinenko}%
  \BibitemOpen
  \bibfield  {author} {\bibinfo {author} {\bibfnamefont {M.}~\bibnamefont
  {Mahmoodi}}, \bibinfo {author} {\bibfnamefont {S.~H.}\ \bibnamefont
  {Tavassoli}}, \bibinfo {author} {\bibfnamefont {O.}~\bibnamefont {Takayama}},
  \bibinfo {author} {\bibfnamefont {J.}~\bibnamefont {Sukham}}, \bibinfo
  {author} {\bibfnamefont {R.}~\bibnamefont {Malureanu}},\ and\ \bibinfo
  {author} {\bibfnamefont {A.~V.}\ \bibnamefont {Lavrinenko}},\ }\bibfield
  {title} {\bibinfo {title} {Existence conditions of high-k modes in finite
  hyperbolic metamaterials},\ }\href {https://doi.org/10.1002/lpor.201800253}
  {\bibfield  {journal} {\bibinfo  {journal} {Laser Photonics Rev.}\ }\textbf
  {\bibinfo {volume} {13}},\ \bibinfo {pages} {1800253} (\bibinfo {year}
  {2019})}\BibitemShut {NoStop}%
\bibitem [{\citenamefont {John}\ \emph {et~al.}(2022)\citenamefont {John},
  \citenamefont {Slassi}, \citenamefont {Sun}, \citenamefont {Sun},
  \citenamefont {Bachelet}, \citenamefont {Pénuelas}, \citenamefont
  {Saint-Girons}, \citenamefont {Orobtchouk}, \citenamefont {Ramanathan},
  \citenamefont {Calzolari},\ and\ \citenamefont {Cueff}}]{hyperwave}%
  \BibitemOpen
  \bibfield  {author} {\bibinfo {author} {\bibfnamefont {J.}~\bibnamefont
  {John}}, \bibinfo {author} {\bibfnamefont {A.}~\bibnamefont {Slassi}},
  \bibinfo {author} {\bibfnamefont {J.}~\bibnamefont {Sun}}, \bibinfo {author}
  {\bibfnamefont {Y.}~\bibnamefont {Sun}}, \bibinfo {author} {\bibfnamefont
  {R.}~\bibnamefont {Bachelet}}, \bibinfo {author} {\bibfnamefont
  {J.}~\bibnamefont {Pénuelas}}, \bibinfo {author} {\bibfnamefont
  {G.}~\bibnamefont {Saint-Girons}}, \bibinfo {author} {\bibfnamefont
  {R.}~\bibnamefont {Orobtchouk}}, \bibinfo {author} {\bibfnamefont
  {S.}~\bibnamefont {Ramanathan}}, \bibinfo {author} {\bibfnamefont
  {A.}~\bibnamefont {Calzolari}},\ and\ \bibinfo {author} {\bibfnamefont
  {S.}~\bibnamefont {Cueff}},\ }\bibfield  {title} {\bibinfo {title} {Tunable
  optical anisotropy in epitaxial phase-change vo2 thin films},\ }\href
  {https://doi.org/doi:10.1515/nanoph-2022-0153} {\bibfield  {journal}
  {\bibinfo  {journal} {Nanophotonics}\ }\textbf {\bibinfo {volume} {11}},\
  \bibinfo {pages} {3913} (\bibinfo {year} {2022})}\BibitemShut {NoStop}%
\end{thebibliography}%

\end{document}